\documentclass[aps,prab,reprint,floatfix,superscriptaddress,amsmath,amssymb]{revtex4-2}

\usepackage{graphicx}
\usepackage{bm}
\usepackage{booktabs}
\usepackage{tikz}
\usetikzlibrary{arrows.meta}
\usepackage[colorlinks=true,linkcolor=blue,citecolor=blue,urlcolor=blue]{hyperref}

\newcommand{\dd}{\mathrm{d}}

\begin{document}

\title{QUARZ: a quasi-static particle-in-cell code with non-uniform radial grids\\ for plasma wakefields}

\author{Alexander Pukhov}
\author{Thomas C. Wilson}
\affiliation{Institut f\"ur Theoretische Physik I, Heinrich-Heine-Universit\"at D\"usseldorf, 40225 D\"usseldorf, Germany}

\date{\today}

\begin{abstract}
We present a quasi-static particle-in-cell scheme for beam-driven plasma wakefields in cylindrical geometry that works on an arbitrary non-uniform radial grid. The field equations are discretized with hat-function finite elements and conservative finite volumes on any monotone set of radial nodes, and the transverse magnetic field is obtained explicitly, without iterations, following Baxevanis and Stupakov. An optional azimuthal mode $m=1$ describes beam offsets, tilts and hosing. The discretization keeps the discrete Gauss law exact, deposits a uniform plasma exactly on every node including the axis, and remains second-order accurate on smoothly stretched grids. Resolution can therefore be placed where the physics needs it. For a tightly focused witness beam that pulls the plasma ions onto the axis, a grid that is fine only near the axis gives the same near-axis fields as a uniform grid with the same axis resolution, with thirty times fewer cells and in thirty times less computing time. Such a grid also resolves the singular closure of a cold-plasma bubble, which a uniform grid hides by smoothing it over one cell; smoothing the plasma sources over a fixed length, chosen independently of the grid, regularizes the closure and makes the wake behind it converge. The scheme is implemented in the open-source C++/Kokkos code QUARZ (Quasistatic Arbitrary-resolution RZ), which runs on CPUs and GPUs and uses MPI with a pipelined decomposition along the co-moving coordinate. Simulations of the self-modulation of the AWAKE proton bunch agree with the three-dimensional code HiPACE++. QUARZ is, to our knowledge, the first production-grade simulation code in plasma physics that is \emph{zero-touch}: all of its source code, tests and documentation were written by an AI agent directed by a physicist who did not write a single line of the code.
\end{abstract}

\maketitle

%---------------------------------------------------------------------------
\section{Introduction}\label{sec:intro}

Quasi-static codes make start-to-end simulations of plasma wakefield accelerators affordable. They exploit the separation of time scales between the plasma response and the evolution of a relativistic driver~\cite{Mora1997}. The plasma is advanced slice by slice in the co-moving variable $\xi = t - z$, and the beam is pushed with a time step set by its betatron period. Established implementations include LCODE~\cite{Lotov2003,Sosedkin2016}, QuickPIC~\cite{Huang2006,An2013}, QPAD~\cite{Li2021}, HiPACE and HiPACE++~\cite{Mehrling2014,Diederichs2022} and Wake-T~\cite{FerranPousa2026}.

Many current problems are strongly multiscale in the radial direction. A low-emittance witness can be focused to a radius of order $10^{-2}\,c/\omega_p$, and it pulls in the plasma ions to densities several times $n_0$ on the axis~\cite{Rosenzweig2005,An2017}. The wake, however, extends over several $c/\omega_p$, and the conducting boundary has to be farther still. A uniform grid must resolve the smallest scale everywhere, so the number of cells, and the cost of each slice, grows with the ratio of the two scales.

Cartesian codes address this with mesh refinement~\cite{Diederichs2022}. In axisymmetric geometry a simpler option is available: a single radial grid with non-uniform spacing, fine on the axis and coarse outside. Ferran Pousa \textit{et al.} recently obtained a similar flexibility in a gridless formulation, by spacing the plasma macro-particles non-uniformly~\cite{FerranPousa2026}. Here we keep a grid, which gives a direct field solve, linear cost in the number of cells, and a natural extension to azimuthal modes.

The contributions of this paper are:
(i) a discretization of the quasi-static field equations on arbitrary radial nodes: hat-function volumes, a lumped-mass finite-element Laplacian for $m=0$ and conservative finite-volume operators for the vector components and for $m=1$ (Sec.~\ref{sec:disc});
(ii) its combination with the explicit, iteration-free $\bm{B}_\perp$ solve of Baxevanis and Stupakov~\cite{Baxevanis2018}, extended to the $m=1$ mode with a short Picard iteration for the mode coupling (Secs.~\ref{sec:model}, \ref{sec:m1});
(iii) exact discrete properties: Gauss law, uniform-plasma deposition including the axis node, and $E_z$ as the exact discrete $\xi$-derivative of $\psi$ (Sec.~\ref{sec:disc});
(iv) a performance-portable implementation with MPI pipelining along $\xi$ (Sec.~\ref{sec:impl}) and a validation set, including a grid study for a pinched witness with ion motion and a benchmark of proton-bunch self-modulation against HiPACE++ (Sec.~\ref{sec:valid}).

QUARZ was also developed in an unusual way. We call it a \emph{zero-touch} code: every line of its source code, of its validation suite and of its documentation was written by an AI agent, the large language model Claude~\cite{Claude} working with a compiler, a shell and the test suite, while one of us (A.P.) directed the work without writing or editing a single line of code, and the other (T.C.W.) tested it on supercomputers. The human contributions were the ones that require physical judgment: the choice of the problem and of the methods, physical conventions (for example, which plasma density the quasi-static box sees), the convergence criteria that decide whether a result can be trusted, and the interpretation of the results. The agent was held to a fixed rule: no feature counts as done before it passes a validation section against an independent analytic or numerical reference, and every change is rerun against the whole suite, so the correctness of the code rests on tests rather than on line-by-line human review. Every commit in the public repository~\cite{QUARZ} records the AI session that produced it, so the claim can be checked. To our knowledge QUARZ is the first production-grade code of this kind in plasma physics; it runs on CPU clusters and GPUs and has been benchmarked independently against HiPACE++ (Sec.~\ref{sec:awake}).

%---------------------------------------------------------------------------
\section{Quasi-static model}\label{sec:model}

We use normalized units: time in $1/\omega_p$, lengths in $c/\omega_p$, densities in $n_0$, momenta in $m_s c$ and fields in $m_e c\,\omega_p/e$. With $\xi = t - z$ and the wake potential $\psi = \phi - A_z$, the transverse force on a particle moving with $v_z = c$ is $\bm{W}_\perp = \bm{E}_\perp + \hat{\bm z}\times\bm{B}_\perp = -\nabla_\perp\psi$. In the quasi-static approximation the fields of one slice follow from the sources of that slice:
\begin{align}
\nabla_\perp^2 \psi &= -(\rho - J_z), \label{eq:psi}\\
\nabla_\perp^2 E_z &= -\partial_\xi(\rho - J_z), \label{eq:ez}\\
\nabla_\perp^2 B_z &= -(\nabla_\perp\times\bm{J}_\perp)_z, \label{eq:bz}\\
\nabla_\perp^2 \bm{B}_\perp &= \hat{\bm z}\times\left(\partial_\xi \bm{J}_\perp + \nabla_\perp J_z\right). \label{eq:bperp}
\end{align}
The term $\partial_\xi\bm{J}_\perp$ is not known on the current slice. Following Baxevanis and Stupakov~\cite{Baxevanis2018}, it is computed from the particle equations of motion. Its part proportional to $\bm{B}_\perp$ is moved to the operator. In complex notation $B_+ = B_x + iB_y$, Eq.~\eqref{eq:bperp} becomes a screened Poisson (Helmholtz) problem,
\begin{align}
\left(\nabla_\perp^2 - \chi\right) B_+ &= i\left[(\partial_x + i\partial_y) J_z + S_+\right], \label{eq:helm}\\
\chi &= \sum_p \frac{q_p^2 w_p}{m_p(\gamma_p - p_{z,p})}\,\delta_p \ge 0, \label{eq:chi}
\end{align}
where $J_+ = J_x + iJ_y$ and
\begin{align}
S_+ &= \sum_p q_p w_p \left[ a_{+,p}\, \delta_p - u_{+,p}\, (\bm{u}_{\perp,p}\cdot\nabla_\perp)\, \delta_p \right], \label{eq:splus}\\
a_+ &= \frac{\hat q}{\Delta^2}\left(\gamma W_+ - i p_+ B_z\right) - \frac{p_+}{\Delta^2}\,\frac{\dd\Delta}{\dd\xi}. \label{eq:aplus}
\end{align}
Here $u_+ = p_+/\Delta = (p_x + i p_y)/\Delta$ is the transverse velocity $\dd x_+/\dd\xi$ of a particle, $\delta_p = \delta(\bm{x}_\perp - \bm{x}_{\perp,p})$ (in the code the hat-function shape of Sec.~\ref{sec:disc}), and $\dd\Delta/\dd\xi$ is given by Eq.~\eqref{eq:ddelta}. $S_+$ is the $\xi$-derivative of $J_+ = \sum_p q_p w_p u_{+,p}\delta_p$ along the particle orbits, $\dd u_+/\dd\xi = a_+ - i\hat q B_+/\Delta$, without its part $-i\chi B_+$: the first term in Eq.~\eqref{eq:splus} is the acceleration of the particles by $\bm{W}_\perp$ and $B_z$ (and, through $\Delta$, by $E_z$), the second is the transport of the current by the moving particles. Since $\chi\ge 0$, the operator is negative definite and the solve needs no iteration. Finally $E_+ = W_+ - iB_+$.

\paragraph{Plasma particles} carry $(x, y, p_x, p_y, \Delta = \gamma - p_z, w)$, with $\Delta$ conserved in a static potential. With $\hat q = q/m$ they obey
\begin{align}
\frac{\dd x}{\dd\xi} &= \frac{p_x}{\Delta}, \label{eq:eom}\\
\frac{\dd p_x}{\dd\xi} &= \hat q\left[\frac{\gamma W_x}{\Delta} + B_y + \frac{p_y B_z}{\Delta}\right], \\
\frac{\dd \Delta}{\dd\xi} &= \hat q\left[\frac{\bm{p}_\perp\cdot\bm{W}_\perp}{\Delta} - E_z\right], \label{eq:ddelta}
\end{align}
and the analogous equations for $y$. The weight $w$ of a plasma particle is the number of real particles per unit length in $\xi$ that it represents (a ring for $m=0$, a ring segment for $m=1$), in units of $n_0 c^2/\omega_p^2$. It is set where the plasma enters the box at rest, $w = n\,\delta A$ with $n$ the local density and $\delta A$ the cross-section of the ring, and it stays constant along the orbit. A particle crosses a slice in the time $1/(1-v_z) = \gamma/\Delta$, so its sources are
\begin{align}
\rho &= \sum_p q_p w_p \frac{\gamma_p}{\Delta_p}\,\delta_p, \quad
\bm{J} = \sum_p q_p w_p \frac{\bm{p}_p}{\Delta_p}\,\delta_p, \nonumber\\
\rho - J_z &= \sum_p q_p w_p\, \delta_p,
\label{eq:plsources}
\end{align}
where $\delta_p$ is the transverse shape of Eq.~\eqref{eq:splus}. They are integrated with an Adams--Bashforth scheme of order 1 to 5 (default 3). Particles are Cartesian in the transverse plane, so they cross the axis without special treatment. Any number of species, including mobile ions, can be used. Particles with $\Delta$ below a threshold are trapped and are removed, since the quasi-static model cannot follow them.

\paragraph{Beam particles} are full 3D macro-particles $(x, y, p_x, p_y, p_z, \xi, w)$, where $w$ is the number of real particles a macro-particle represents, in units of $n_0 c^3/\omega_p^3$; they are deposited with a shape in $r$ and in $\xi$, $\rho_b = \sum_p q_p w_p\, \delta_p\,\delta(\xi - \xi_p)$ and $\bm{J}_b = \sum_p q_p w_p \bm{v}_p\, \delta_p\,\delta(\xi - \xi_p)$. They are advanced in lab time $t$ with a time step $\Delta t \gg \Delta\xi$. We use the Vay pusher~\cite{Vay2008} by default, because the beam's own fields nearly cancel ($E_r\approx B_\theta$) and the Boris pusher does not preserve this balance; the Higuera--Cary pusher~\cite{Higuera2017} and an implicit scheme with radiation reaction are available. The beam charge and current, $(\rho - J_z)_b$, enter the right-hand sides above.

%---------------------------------------------------------------------------
\section{Discretization on an arbitrary radial grid}\label{sec:disc}

The radial nodes $0 = r_0 < r_1 < \dots < r_N = R$ are arbitrary; $h_j = r_{j+1} - r_j$ and $r_{j+1/2} = (r_j + r_{j+1})/2$. All fields and sources live on the nodes. In practice the grid is built from regions with prescribed spacings, joined geometrically so that the ratio of neighbouring cells stays below a bound (typically 1.05--1.1).

\paragraph{Hat-function volumes.} Charge assignment and field interpolation use the linear hat functions $W_j(r)$ of the non-uniform cells. The volume associated with node $j$ is the exact integral of its hat function,
\begin{align}
V_j &= 2\pi\!\int W_j(r)\,r\,\dd r \nonumber\\
    &= 2\pi\left[\frac{r_{j-1}h_{j-1}}{2} + \frac{h_{j-1}^2}{3} + \frac{r_j h_j}{2} + \frac{h_j^2}{6}\right], \label{eq:vol}
\end{align}
and a density on node $j$ is $\sum_p w_p W_j(r_p)/V_j$. This generalizes Verboncoeur's volume correction~\cite{Verboncoeur2001} to non-uniform cells. Because the volumes are exact, a ring placed at the volume centroid of each radial sub-cell deposits a uniform density exactly on every node, including $r=0$. We load the plasma this way; placing rings at the mid-points instead gives density errors of order 10\% on the axis.

\paragraph{$m=0$ Laplacian.} We use lumped-mass linear finite elements, which with the same $V_j$ coincide with the conservative flux form:
\begin{equation}
(L_0 f)_j = \frac{2\pi}{V_j}\left[r_{j+\frac12}\frac{f_{j+1}-f_j}{h_j} - r_{j-\frac12}\frac{f_j - f_{j-1}}{h_{j-1}}\right]. \label{eq:L0}
\end{equation}
The sum $\sum_j V_j (L_0 f)_j$ telescopes to the flux through the wall, so the discrete Gauss law holds exactly. The axis row contains only the outward flux and needs no regularity condition.

\paragraph{Higher-order operators.} The vector components ($B_+$ modes) and the $m=1$ scalars need $L_n f = (1/r)(r f')' - n^2 f/r^2$ with $n = 1, 2$. We write it in the factorized form $L_n f = r^{n-1}\partial_r\!\left[r^{1-2n}\partial_r(r^n f)\right]$ and discretize it with conservative finite volumes:
\begin{align}
(L_n f)_j &= \frac{2\,r_j^{\,n-1}}{h_{j-1}+h_j}\left(D_{j+1/2} - D_{j-1/2}\right), \label{eq:Ln}\\
D_{j+1/2} &= \frac{r_{j+1}^{\,n} f_{j+1} - r_j^{\,n} f_j}{h_j\, r_{j+1/2}^{\,2n-1}}, \nonumber
\end{align}
with $f(0)=0$. For $n=1$ this is the familiar $\partial_r[(1/r)\partial_r(rB)]$. The factorized form keeps the stencil tridiagonal and well conditioned near the axis on any grid. Each slice therefore requires only tridiagonal solves: Thomas on CPUs and parallel cyclic reduction inside one thread block on GPUs. The $\chi$ term enters the diagonal.

\paragraph{Boundary conditions.} At $r=R$ we impose $\psi = E_z = B_z = 0$ and Dirichlet conditions on $B_r$ and on the $m=1$ parts of $B_+$. For $B_\theta$ we impose the flux condition $(1/r)\partial_r(rB_\theta) = J_z$ at the wall, i.e.\ Amp\`ere's law evaluated at the boundary.

\paragraph{Exact $\xi$-derivatives of the deposits.} The source $\partial_\xi(\rho - J_z)$ of $E_z$, and $\partial_\xi\bm{J}_\perp$, are not differenced between slices. Each particle deposits the exact $\xi$-derivative of its own discrete deposit, $w\,W_j'(r_p)\,\dd r_p/\dd\xi$, plus the azimuthal part for $m=1$. With this choice, $E_z$ is the exact discrete $\xi$-derivative of $\psi$ in every mode.

\paragraph{Gradients} ($\bm{W}_\perp = -\nabla_\perp\psi$ and the Lorentz-force terms) use the three-point second-order formula for non-uniform spacing. On smoothly stretched grids all operators converge at second order; measured orders against manufactured solutions are 1.9--2.0 for $L_0$, $L_1$ and $L_2$ ($h_0 = 0.01\to0.00125$, ratio $\le 1.05$). Abrupt jumps in spacing reduce the accuracy locally to first order. A fine axial mesh also resolves the singular closure of a cold-plasma bubble; Sec.~\ref{sec:closure} shows how it is regularized independently of the grid.

%---------------------------------------------------------------------------
\section{Azimuthal mode $m=1$}\label{sec:m1}

Every scalar is expanded as $f(r,\theta) = f_0(r) + 2\,\mathrm{Re}[f_1(r)e^{i\theta}]$, as in QPAD~\cite{Li2021} and in quasi-cylindrical electromagnetic codes~\cite{Lifschitz2009}. For transverse vectors it is convenient to expand the complex combination $V_+ = V_x + iV_y$:
\begin{equation}
V_+ = v_0(r) + v_1(r)e^{i\theta} + v_2(r)e^{2i\theta}. \label{eq:vplus}
\end{equation}
Here $v_1 = V_{r0} + iV_{\theta0}$ is the axisymmetric part, while $v_0$ and $v_2$ carry $m=1$. Each coefficient $v_k$ obeys a scalar radial equation with the operator $L_k$ of Sec.~\ref{sec:disc}, so the scalar fields ($\psi$, $E_z$, $B_z$) need $L_0$ and $L_1$, and $B_+$ needs $L_0$, $L_1$ and $L_2$. Regularity on the axis gives $v_1(0) = v_2(0) = 0$.

The only new difficulty is the term $\chi B_+$. With $\chi = \chi_0 + 2\,\mathrm{Re}[\chi_1 e^{i\theta}]$, the product couples $b_k$ to $b_{k\pm1}$. We keep $\chi_0$ in the operator and move the $\chi_1$ products to the right-hand side. A few Picard iterations then converge quickly, because $\chi_1$ is small whenever the $m=1$ description itself is valid; three iterations reached 0.3\% in our tests. For a centred beam the $m=1$ part vanishes identically and the $m=0$ result is recovered to round-off.

Plasma particles stay Cartesian. With $m=1$ every loaded ring is split into $n_\theta$ particles at equidistant azimuths (default 8), and each particle deposits with weights $e^{-ik\theta}$. Beam particles gather the fields as mode sums at their own azimuth. To keep sampling noise from seeding a spurious $m=1$ component, beams can be loaded with an $n$-fold rotational quiet start, or deposited analytically: for a displaced Gaussian the $m=0$ and $m=1$ parts are known in closed form in terms of modified Bessel functions $I_0$, $I_1$.

The expansion is truncated at $m=1$. It therefore describes offsets and tilts that are small compared with the beam and bubble radii, which covers the onset of hosing and beam--plasma misalignment.

%---------------------------------------------------------------------------
\section{Implementation and parallelization}\label{sec:impl}

QUARZ is written in C++17 on Kokkos~\cite{Trott2022}, so one source compiles for OpenMP, CUDA, HIP and SYCL. Particles are stored as structures of arrays, and all particle loops are device kernels with atomic deposition. With $m=0$ a slice costs about 12 kernels and 5 tridiagonal solves; with $m=1$ and three Picard iterations, about 20 kernels and 25 solves. All arithmetic is in double precision. On CPUs the tridiagonal systems are solved with the Thomas algorithm. On GPUs we use the partition method of L\'aszl\'o \textit{et al.}~\cite{Laszlo2016}: the $M$ rows are cut into $P\le 256$ chunks of at least 8 rows, each GPU thread eliminates the interior of one chunk, the $2P$ unknowns at the chunk ends form a tridiagonal system that is solved by parallel cyclic reduction in shared memory, and the interior is then recovered. Plain cyclic reduction of all $M$ rows in one thread block takes $\log_2 M$ passes through global memory; on a grid of 3841 nodes it was 4.3 times slower for the whole run (Sec.~\ref{sec:perf}). Host--device traffic is limited to loading the fresh plasma slice and to output. The source code of QUARZ, the input decks of all simulations in this paper and the validation suite are available at \url{https://github.com/AlexMPukhov/QUARZ}~\cite{QUARZ}. QUARZ is a zero-touch code (Sec.~\ref{sec:intro}): its source code and validation scripts, and a first draft of this manuscript, were written by the AI agent Claude~\cite{Claude} under the direction of the authors (see Acknowledgments).

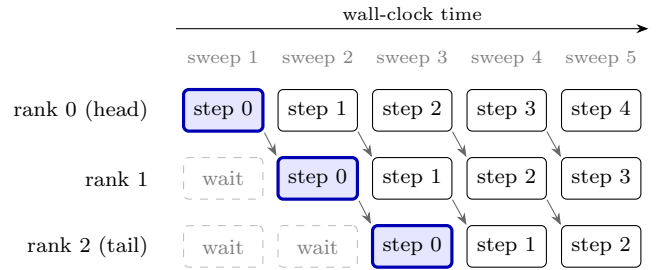
\begin{figure}[t]
\centering
\begin{tikzpicture}[x=1.25cm,y=0.9cm,
  cellbox/.style={draw,rounded corners=2pt,minimum width=1.05cm,minimum height=0.55cm,font=\footnotesize,inner sep=0pt},
  hot/.style={cellbox,very thick,draw=blue!70!black,fill=blue!10},
  idle/.style={cellbox,draw=gray!50,dashed,text=gray},
  msg/.style={-{Stealth[length=1.6mm]},gray!80!black}]
  \foreach \c in {0,...,4} { \node[font=\scriptsize,gray] at (\c,0.75) {sweep \the\numexpr\c+1\relax}; }
  \node[font=\footnotesize,anchor=east] at (-0.7,0)  {rank 0 (head)};
  \node[font=\footnotesize,anchor=east] at (-0.7,-1) {rank 1};
  \node[font=\footnotesize,anchor=east] at (-0.7,-2) {rank 2 (tail)};
  % rank 0
  \node[hot] (a0) at (0,0) {step 0};
  \foreach \c in {1,...,4} { \node[cellbox] (a\c) at (\c,0) {step \c}; }
  % rank 1
  \node[idle] at (0,-1) {wait};
  \node[hot] (b1) at (1,-1) {step 0};
  \foreach \c/\s in {2/1,3/2,4/3} { \node[cellbox] (b\c) at (\c,-1) {step \s}; }
  % rank 2
  \node[idle] at (0,-2) {wait};
  \node[idle] at (1,-2) {wait};
  \node[hot] (c2) at (2,-2) {step 0};
  \foreach \c/\s in {3/1,4/2} { \node[cellbox] (c\c) at (\c,-2) {step \s}; }
  % messages
  \foreach \c/\d in {0/1,1/2,2/3,3/4} { \draw[msg] (a\c.south east) -- (b\d.north west); }
  \foreach \c/\d in {1/2,2/3,3/4} { \draw[msg] (b\c.south east) -- (c\d.north west); }
  \draw[-{Stealth[length=1.6mm]}] (-0.5,1.2) -- (4.5,1.2) node[midway,above,font=\scriptsize] {wall-clock time};
\end{tikzpicture}
\caption{Time pipelining along $\xi$ with three ranks. Arrows: one message per time step from rank $r$ to $r+1$ (plasma particles with their Adams--Bashforth history, $B_+$ and $(\rho-J_z)_b$ of the last slice, and beam particles that slipped downstream). Highlighted: one time step crossing the box. After $P-1$ sweeps all ranks are busy.}
\label{fig:pipeline}
\end{figure}

\paragraph{MPI pipelining along $\xi$.} The box is split into $P$ contiguous blocks of slices, one per rank, with rank 0 at the head. The plasma enters at the head, so rank $r$ can sweep step $n$ only after rank $r-1$ has finished it; but rank $r-1$ can then start step $n+1$ (Fig.~\ref{fig:pipeline}). This time pipelining, used in QuickPIC~\cite{Feng2009} and HiPACE++~\cite{Diederichs2022}, gives a speed-up of $PN/(N+P-1)$ for $N$ sweeps. Once per step, rank $r$ sends rank $r+1$ a single message with: the plasma particles leaving its last slice, with their Adams--Bashforth history; $B_+$ of the last slice, used as the starting guess of the next solve; $(\rho - J_z)_b$ of the last slice, needed for the $\xi$-derivative of the beam source; and beam particles that slipped into the next block during the push ($\dd z/\dd t\le c$, so beam particles only move towards larger $\xi$).

For the pipeline, the beam deposit and the field gather of a slice must not involve slices of another rank at another time level. Beam particles therefore use nearest-slice assignment in $\xi$, with the full hat-function shape in $r$. A serial run with the same assignment performs identical arithmetic, so serial and parallel runs can be compared directly (Sec.~\ref{sec:valid}). On GPUs each rank drives one device and messages pass through host buffers, so no CUDA-aware MPI is needed; the message is small compared with the work of a sweep.

%---------------------------------------------------------------------------
\section{Validation and benchmarks}\label{sec:valid}

The stretched grid reproduces every reference to the same accuracy as a fine uniform grid, and it is markedly more accurate per cell for near-axis structure. Unless stated otherwise, the runs used the OpenMP backend on 2 cores.

\subsection{Verification}

Table~\ref{tab:verif} summarizes the verification tests.

\begin{table*}[t]
\caption{Verification tests (stretched grids have neighbour ratio $\le1.05$).}
\label{tab:verif}
\begin{ruledtabular}
\begin{tabular}{p{0.42\textwidth}p{0.52\textwidth}}
Test & Result \\ \hline
$L_0$, $L_1$, $L_2$ vs manufactured solutions, stretched grids & second order (measured 1.98--2.0) \\
Uniform plasma deposit, stretched grid & exact to $5\times10^{-13}$, including $r=0$ \\
Linear wake ($n_b=10^{-3}$) vs Green's function in a conducting cylinder & $E_z(0)$ error 0.13\% (uniform, 0.01) and 0.12\% (stretched $0.002\to0.05$, 389 cells); the error scales with $n_b$, i.e.\ it is the physical nonlinearity \\
Blowout ($n_b=10$, $\sigma_r=0.3$) & $(E_r-B_\theta)/r = 0.5000$ in the bubble; $E_z$ converged to 4 digits for $\Delta r,\Delta\xi = 0.01\to0.0025$; stretched grid within 0.08\% \\
Betatron oscillation in an ion channel, $\gamma=1000$ & $r_\mathrm{rms}$ follows $r_0|\cos(t/\sqrt{2\gamma})|$ to 0.26\% \\
Witness energy gain, rigid driver & $\dd\langle\gamma\rangle/\dd t = 0.4477$ vs $-E_z = 0.4471$ from the wake \\
Matched witness ($\sigma_r=0.01$, $\varepsilon_n=0.01$, $\gamma=2\times10^4$), $400/\omega_p$ & $\varepsilon_n$ preserved ($0.0097\to0.0097$) for $\Delta t = 10$ and 20; Boris: 9\% spurious growth \\
$m=1$: displaced driver ($d=0.01$, linear regime) vs $-d\,\partial_r$ of the $m=0$ solution & L2 errors $\psi$ 0.08\%, $E_z$ 0.4\%, $E_r$ 0.3\%, $B_\theta$ 0.19\%; sine parts $\le 6\times10^{-14}$ \\
$m=1$: offset beam ($x_0=0.05$) in an ion channel & centroid follows $x_0\cos(t/\sqrt{2\gamma})$ to 0.26\% \\
\end{tabular}
\end{ruledtabular}
\end{table*}

\subsection{Pinched witness with ion motion}\label{sec:pinch}

\begin{figure*}[t]
\centering
\includegraphics[width=\textwidth]{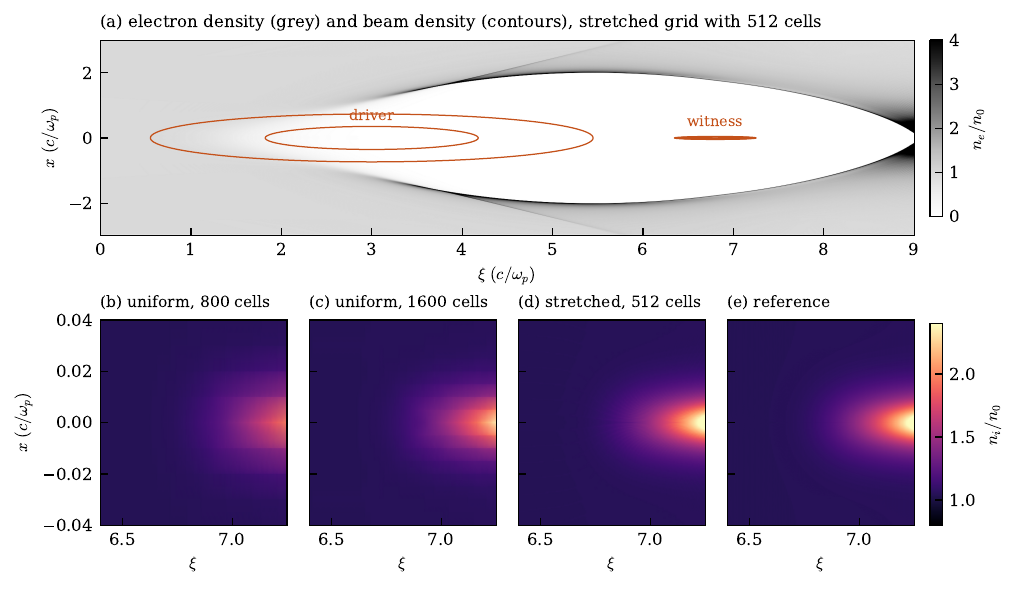}
\caption{Pinched witness with mobile hydrogen ions (single quasi-static solve, noise-free beams). (a) Electron density and beam density contours over the whole box, stretched grid with 512 cells. (b)--(e) Ion density near the axis inside the witness: uniform grids with $\Delta r = 0.01$ (800 cells) and $0.005$ (1600 cells), the stretched grid of (a) with $h_0 = 5\times10^{-4}$ on the axis, and the reference ($h_0 = 1.6\times10^{-5}$, 14\,617 cells). On the uniform grids the ion spike is spread over the first cells and its peak is too low; the stretched grid is indistinguishable from the reference.}
\label{fig:maps}
\end{figure*}

This is the case the non-uniform grid is designed for. A Gaussian driver ($n_b=10$, $\sigma_r=0.3$, $\sigma_\xi=1$) drives a blowout in a hydrogen plasma column of radius 7 inside a conducting wall at $R=8$. A witness with peak density $1.25\times10^4$, $\sigma_r=0.01$ and $\sigma_\xi=0.1$ (matched for $\varepsilon_n=0.01$ at $\gamma=2\times10^4$) sits at $\xi=6.8$ and pulls the ions onto the axis (Fig.~\ref{fig:maps}). Within the witness the ion density on the axis rises to $2.3\,n_0$, and the focusing gradient at $r=0.01$ is 0.88 instead of the 0.5 of an unperturbed ion column.

To separate grid errors from sampling noise, both beams are deposited analytically in this test (Sec.~\ref{sec:m1}), and we compare one quasi-static solve on a sequence of grids. The uniform family has $\Delta r = 0.02\ldots0.00125$ (400 to 6400 cells). The stretched family scales all three spacings of the 512-cell grid ($5\times10^{-4}$ for $r<0.05$, $0.01$ for $r<2.5$, $0.05$ outside, ratio $\le 1.05$) by factors 4 to 1/8 (172 to 3697 cells). The reference is the same stretched family at factor 1/32 (14\,617 cells). It agrees with a uniform grid of 32\,000 cells ($\Delta r = 2.5\times10^{-4}$) to $2\times10^{-4}$, the accuracy of that uniform grid. All runs use $\Delta\xi = 0.005$. Errors are maxima over the witness ($6.4<\xi<7.2$), relative to the maximum of the reference.

\begin{figure*}[t]
\centering
\includegraphics[width=\textwidth]{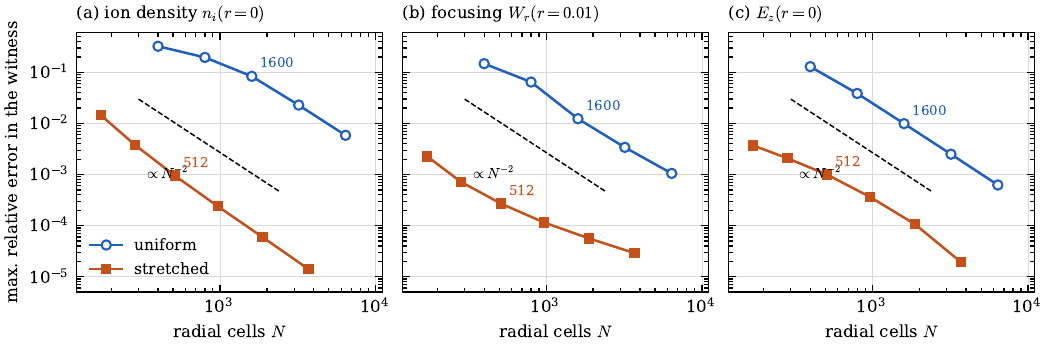}
\caption{Error versus number of radial cells for the pinched witness of Fig.~\ref{fig:maps}: (a) ion density on the axis, (b) focusing field $W_r = E_r - B_\theta$ at $r=0.01$ (the witness radius), (c) $E_z$ on the axis. Open circles: uniform grids; squares: stretched grids with the same shape and all spacings scaled. The grids of Table~\ref{tab:pinch} are labelled. The flattening of the finest stretched points in (b) reflects the accuracy of the reference.}
\label{fig:convergence}
\end{figure*}

The point of the stretched grid is that it keeps the resolution where it matters and drops it elsewhere. A uniform grid with the axis cell of the 512-cell stretched grid, $\Delta r = 5\times10^{-4}$, needs 16\,000 cells. Both grids give the same answer: the ion density on the axis and the focusing field at $r = 0.01$ and $0.03$ agree to $1$--$4\times10^{-5}$ in the witness, and both agree with the reference to $3\times10^{-4}$ in $W_r$ (Table~\ref{tab:pinch}). Only $E_z(0)$ differs, by $10^{-3}$: it is set by the bubble sheath, which the stretched grid resolves with cells of 0.01. The uniform grid takes 36.6~s per sweep, the stretched grid 1.2~s on the same two cores: the same result 30 times faster, in proportion to the number of cells.

Figure~\ref{fig:convergence} shows that both families converge at second order, as expected from Sec.~\ref{sec:disc}, but the stretched curves lie one to two orders of magnitude lower. To reach $10^{-3}$ in the focusing field, the uniform grid needs 6400 cells (15.2~s per sweep), the stretched grid fewer than 300 (0.9~s per sweep): 20 times fewer cells and 17 times less time. Table~\ref{tab:pinch} lists the grids of Fig.~\ref{fig:maps}. With a third of the cells and a third of the time of the 1600-cell uniform grid, the 512-cell stretched grid is 10 times more accurate in $E_z$ and 45--90 times more accurate in the ion density and the focusing field.

\begin{table*}[t]
\caption{Pinched witness with mobile hydrogen ions: maximum relative errors in the witness with respect to the reference of Fig.~\ref{fig:convergence}. $W_r = E_r - B_\theta$. Times per sweep of 1461 slices on 2 CPU cores.}
\label{tab:pinch}
\begin{ruledtabular}
\begin{tabular}{lrrrrrr}
Grid & cells & time per sweep (s) & $n_i(0)$ & $W_r(0.01)$ & $W_r(0.03)$ & $E_z(0)$ \\ \hline
uniform 0.01 & 800 & 2.0 & 20\% & 6.5\% & 3.0\% & 3.9\% \\
uniform 0.005 & 1600 & 4.1 & 8.4\% & 1.2\% & 0.83\% & 0.99\% \\
stretched $0.0005\to0.01\to0.05$ & 512 & 1.4 & 0.10\% & 0.027\% & 0.015\% & 0.10\% \\
uniform 0.0005 & 16\,000 & 36.6 & 0.094\% & 0.028\% & 0.015\% & 0.011\% \\
\end{tabular}
\end{ruledtabular}
\end{table*}

With randomly sampled macro-particle beams ($4\times10^5$ driver and $2\times10^5$ witness particles) the errors of the 512-cell stretched grid rise to 0.3\% in $W_r(0.01)$ and 5\% in $n_i(0)$. They are then set by beam sampling noise, not by the grid: fine cells near the axis must be paired with enough beam particles there, or with a quiet start.

\begin{figure*}[t]
\centering
\includegraphics[width=\textwidth]{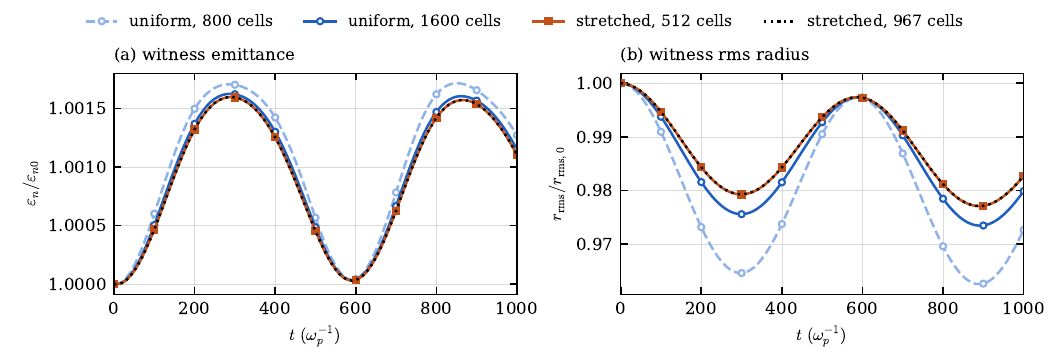}
\caption{Evolution of the pinched witness of Fig.~\ref{fig:maps} behind a rigid driver, with mobile hydrogen ions, on two uniform and two stretched grids ($\Delta t = 10$, $2\times10^5$ witness particles). (a) Normalized emittance, (b) rms radius, both relative to their initial values. The 512-cell stretched grid coincides with the 967-cell one; the uniform grids overestimate the envelope oscillation.}
\label{fig:evolution}
\end{figure*}

\paragraph*{Witness evolution.} The field errors carry over into the beam dynamics. Figure~\ref{fig:evolution} follows the witness ($2\times10^5$ macro-particles) for 100 steps of $\Delta t = 10$, behind a rigid, analytically deposited driver, so that only the witness, the plasma electrons and the ions evolve. The witness was matched to the unperturbed ion column; the collapsed ions focus it more strongly, so its rms radius oscillates. On the 800- and 1600-cell uniform grids the first minimum of $r_\mathrm{rms}$ is 3.7\% and 2.7\% below the initial value, i.e.\ the amplitude is overestimated by 64\% and 16\%. The 512-cell stretched grid gives 2.29\%, within 0.1\% of the amplitude on a stretched grid twice as fine (967 cells); its sweep is 3 times faster than that of the 1600-cell uniform grid. The emittance oscillates by only 0.16\% in all runs. The energy gain over the run, $\Delta\gamma\approx 329$, is 1.2\% and 0.3\% too low on the uniform grids and within 0.07\% on the 512-cell stretched grid.

\subsection{Bubble closure on a fine axial mesh}\label{sec:closure}

The fine axial mesh that resolves the witness also resolves the rear of the bubble, where the sheath electrons return to the axis. For a cold plasma and an exactly axisymmetric driver they all cross the axis within a short interval of $\xi$. Their trajectories form a caustic, and the on-axis electron density and $E_z$ are singular there~\cite{Lotov2003}. The spike carries no energy, and it does not exist in a real plasma, which is warm and never perfectly symmetric. What matters is the wake behind it.

\begin{figure*}[t]
\centering
\includegraphics[width=\textwidth]{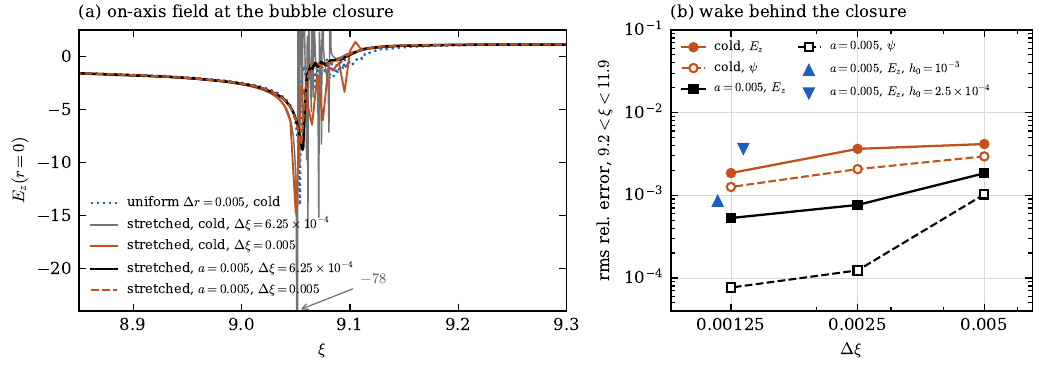}
\caption{Bubble closure in the pinched-witness case of Fig.~\ref{fig:maps}, with the box extended to $\xi=12$ (single quasi-static solve). (a) $E_z$ on the axis at the closure: uniform grid $\Delta r = 0.005$; stretched grid ($h_0 = 5\times10^{-4}$) with a cold plasma and with smoothing of the plasma sources, Eq.~\eqref{eq:smooth}, $a = 0.005$. The cold spike reaches $-78$ at $\Delta\xi = 6.25\times10^{-4}$. (b) Rms relative error of $E_z(0)$ and $\psi(0)$ behind the closure ($9.2<\xi<11.9$) versus $\Delta\xi$, with respect to $\Delta\xi = 6.25\times10^{-4}$ of the same model. Triangles: $a = 0.005$ with axis cells $10^{-3}$ and $2.5\times10^{-4}$ ($\Delta\xi = 0.00125$).}
\label{fig:closure}
\end{figure*}

To study the closure, we extend the box of Sec.~\ref{sec:pinch} to $\xi = 12$; the bubble closes at $\xi \approx 9.05$ (Fig.~\ref{fig:closure}). A uniform grid regularizes the caustic implicitly, over one cell. With $\Delta r = 0.005$ the peak field is $E_z(0) \approx -14$ and the on-axis density about $1.7\times10^4\,n_0$. The stretched grid with $h_0 = 5\times10^{-4}$ removes this hidden smoothing. As $\Delta\xi$ decreases from 0.005 to $6.25\times10^{-4}$, the peak grows from $-15$ to $-78$ and the on-axis density from $7.5\times10^4$ to about $10^6\,n_0$. The wake behind the closure ($9.2 < \xi < 11.9$) still converges, but slowly and irregularly: the rms error of $E_z(0)$ is 0.42\%, 0.36\% and 0.19\% for $\Delta\xi = 0.005$, 0.0025 and 0.00125 [Fig.~\ref{fig:closure}(b)].

Rather than let the grid set the regularization, we make it an explicit parameter, a length $a$ independent of the mesh. In each slice, all plasma sources ($\rho - J_z$ and its $\xi$-derivative, $J_z$, $\bm{J}_\perp$, $\chi$, the explicit part of $\partial_\xi\bm{J}_\perp$) and the neutralizing ion background are replaced by
\begin{equation}
\tilde f = \left(1 - a^2\nabla_\perp^2\right)^{-1} f , \label{eq:smooth}
\end{equation}
so that every ring acquires the radial profile of the two-dimensional Green's function $K_0(|\bm{r}-\bm{r}'|/a)/(2\pi a^2)$. Each azimuthal component of angular number $n$ is filtered with its own operator, $(1 - a^2 L_n)\tilde f = f$, using Eqs.~\eqref{eq:L0} and \eqref{eq:Ln} with a zero-flux wall for $n=0$. This adds one tridiagonal solve per source and slice. The filter conserves the total charge exactly, leaves a uniform plasma unchanged, and is the same in every slice, so $E_z$ remains the exact discrete $\xi$-derivative of $\psi$. Beams are not filtered.

With $a = 0.005$ (ten axis cells) the closure is resolved: the peak field is $-8.3$ to $-8.8$ and the on-axis density $1$--$2\times10^4\,n_0$, for all $\Delta\xi$ and for axis cells from $10^{-3}$ to $2.5\times10^{-4}$. The wake behind the closure now converges cleanly, with errors of 0.18\%, 0.08\% and 0.05\% in $E_z(0)$ (0.10\%, 0.012\% and 0.008\% in $\psi$). Changing the axis cell to $10^{-3}$ or $2.5\times10^{-4}$ changes it by 0.09\% and 0.37\%. The price is a change of the model of order $a$: compared with the cold solution, the wake behind the closure differs by 0.4\% in $E_z(0)$ and 0.6\% in $\psi(0)$, and the field in the witness by $2\times10^{-4}$, far below the errors of the uniform grids in Table~\ref{tab:pinch}. The results of Sec.~\ref{sec:pinch}, computed without smoothing, therefore stand. The filter costs 10--25\% of the sweep time.

\subsection{Self-modulation of a long proton bunch}\label{sec:awake}

As a benchmark at experimental scale, we simulate the seeded self-modulation of the proton bunch in the first plasma cell of AWAKE Run~2, following the code comparison of Farmer \textit{et al.}~\cite{Farmer2026}. The plasma has the density $n_0 = 7\times10^{14}\,\mathrm{cm^{-3}}$ ($k_p^{-1} = 200\,\mu$m, $E_0 = mc\omega_p/e = 2.5$~GV/m), radius $5\,k_p^{-1}$ (1~mm) and immobile ions. A density step of $+3.5\%$ at $z = 1.75$~m stabilizes the wakefield. The bunch contains $3\times10^{11}$ protons of 400~GeV ($\gamma = 427.3$, relative energy spread $3.5\times10^{-4}$) with normalized emittance $2.2\,\mu$m. It is a Gaussian with $\sigma_r = k_p^{-1}$ and $\sigma_z = 253.7\,k_p^{-1}$ (5.1~cm, 170~ps), cut at one $\sigma_z$ ahead of its centre; the sharp front seeds the self-modulation, as the relativistic ionization front does in the experiment. Its peak density is $n_b \approx 0.014\,n_0$.

The box is $1000\,k_p^{-1}$ long with $\Delta\xi = 1/64$ (64\,064 slices), the conducting wall is at $R = 30\,k_p^{-1}$, and the bunch is represented by $2\times10^7$ macro-particles, pushed with $\Delta t = 200\,\omega_p^{-1}$. The plasma is loaded with 128 rings per cell. We use two radial grids: a uniform one with $\Delta r = 1/128$ (3840 cells), and a stretched one with spacings 0.005 for $r<0.1$, 0.01 for $r<1$ and 0.02 up to the plasma edge, growing to 1 at the wall with a neighbour ratio $\le 1.05$ (403 cells). Both are run on CPUs and on GPUs.

\begin{figure}[t]
\centering
\includegraphics[width=\columnwidth]{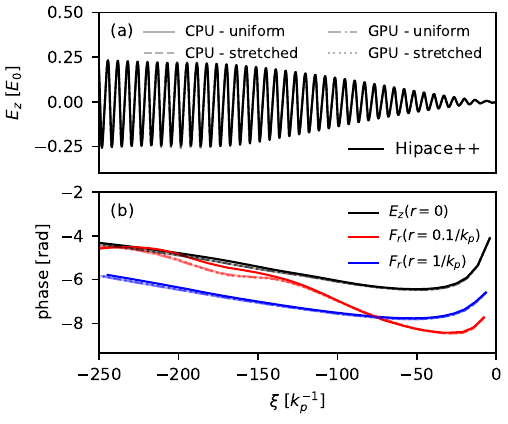}
\caption{Self-modulation of the AWAKE proton bunch after 6~m of plasma: QUARZ on uniform (3840 cells) and stretched (403 cells) radial grids, on CPUs and GPUs, compared with HiPACE++. (a) $E_z$ on the axis, with the head of the bunch at $\xi = 0$. (b) Phase of $E_z(r=0)$ and of the transverse force $F_r = E_r - B_\theta$ at $r = 0.1$ and $1\,k_p^{-1}$, from the zero crossings of the fields. Saturated lines: HiPACE++; light lines: the four QUARZ runs.}
\label{fig:awake}
\end{figure}

Figure~\ref{fig:awake} compares the fields after 6~m of propagation with HiPACE++~\cite{Diederichs2022} (version 24.08). The HiPACE++ run uses a uniform Cartesian grid with $\Delta x = \Delta y = 1/128$, i.e.\ the transverse cell of our uniform grid, and the same $\Delta\xi$, in a box of $16\times16\,k_p^{-2}$, with $5\times5$ plasma particles per cell, $1.2\times10^8$ beam macro-particles and immobile ions. The on-axis $E_z$ grows along the bunch to $0.23\,E_0$ and agrees with HiPACE++ over the whole bunch. Panel (b) shows the phase of three fields, obtained from their zero crossings relative to a harmonic wave~\cite{Farmer2026}: $E_z$ on the axis and the transverse force $F_r = E_r - B_\theta$ at $r = 0.1$ and $1\,k_p^{-1}$. For $E_z(0)$ and $F_r(1)$ the phases of QUARZ and HiPACE++ agree to about 0.1~rad over 250~$k_p^{-1}$, i.e.\ about 40 wavelengths. $F_r$ close to the axis is the most sensitive diagnostic: there the front of the bunch is torn apart by the nonlinear near-axis focusing~\cite{Farmer2026}. Its phase follows HiPACE++ for $\xi > -130$, but differs by up to 0.4~rad between $\xi = -210$ and $-130$. The four QUARZ runs agree among one another to the width of the lines, although their radial cells at $r = 0.1$ differ by a factor of 1.6 and they run on different hardware; and the uniform QUARZ grid has the transverse cell of HiPACE++. The difference is therefore not a matter of resolution. It more likely reflects the different description of the near-axis protons: in three dimensions their sampling noise is not axisymmetric and acts on the tearing of the bunch front, which an $m=0$ model excludes.

\subsection{Parallel performance}\label{sec:perf}

MPI runs were compared file by file with serial runs that use the same nearest-slice beam assignment. Runs without beam particles crossing rank boundaries are bit-identical (witness gain, ion motion, parsed profiles, analytic offset driver with $m=1$). When particles move between ranks, their order in memory changes and the results differ at round-off level: $10^{-10}$ in the fields for hosing ($m=1$, 12 steps) and $5\times10^{-8}$ for a $\gamma=5$ beam slipping through all ranks. On 2 CPU cores the speed-up is 1.77--1.83 for 13--25 sweeps, against 1.86--1.92 for an ideal pipeline.

Larger runs were timed with the AWAKE case of Sec.~\ref{sec:awake} on the uniform grid, without output, on the supercomputers JUWELS (nodes with two 24-core Intel Xeon Platinum 8168 CPUs) and JUPITER (nodes with four NVIDIA GH200 superchips, of which only the GPUs are used). Table~\ref{tab:perf} gives the wall-clock time of one sweep through the box, per rank, i.e.\ the work of a full sweep divided among the ranks. With one MPI rank per core, a JUWELS node is 43 times faster than a single core, an efficiency of 90\%. On GPUs the scaling is ideal up to 16 devices, the largest number tested. One GPU does a sweep in 25~s, 60 times faster than one CPU core and 1.4 times faster than a full 48-core node. Threading within a rank is less efficient: 24 OpenMP threads are only 5.5 times faster than one, so on CPUs we run one MPI rank per core.

\begin{table}[t]
\caption{Wall-clock time of one sweep per rank (s), AWAKE case of Sec.~\ref{sec:awake} on the uniform grid (3840 cells, 64\,064 slices, $2\times10^7$ beam particles). Averages over 50 sweeps (10 on one core). CPU: JUWELS (Intel Xeon Platinum 8168, 48 cores per node); GPU: NVIDIA GH200 (JUPITER).}
\label{tab:perf}
\begin{ruledtabular}
\begin{tabular}{lrrrrr}
CPU cores & 1 & 6 & 12 & 24 & 48 \\
MPI ranks & 1519 & 246 & 125 & 64 & 35 \\
OpenMP threads & 1519 & 444 & 327 & 277 & 83\footnotemark[1] \\ \hline
GPUs & 1 & 2 & 4 & 8 & 16 \\
MPI ranks & 25 & 13 & 6.5 & 3.4 & 1.8 \\
\end{tabular}
\end{ruledtabular}
\footnotetext[1]{2 MPI ranks with 24 threads each, one per socket.}
\end{table}

The full runs of Fig.~\ref{fig:awake} (301 sweeps up to 12~m, with field output) took 667~s on the uniform and 319~s on the stretched grid with 16 GPUs, and 3900~s and 1530~s on 384 CPU cores (8 JUWELS nodes). The stretched grid has ten times fewer cells, but the run is only 2--2.5 times faster: it has half as many plasma particles, and these dominate its cost. With 384 ranks each rank holds only 167 slices, and the pipeline fill limits the efficiency to $N/(N+P-1) = 44\%$ for $N = 301$ sweeps; 16 GPUs work at 95\%.

%---------------------------------------------------------------------------
\section{Conclusions}

A quasi-static RZ scheme can run on any monotone radial grid without losing the properties that make uniform-grid codes robust: exact discrete Gauss law, exact uniform-plasma deposition on the axis, second-order accuracy and an iteration-free $\bm{B}_\perp$ solve. The cost per slice stays linear in the number of cells, so resolution can be spent where the physics needs it. For a pinched witness with ion motion, a stretched grid with the same axis cell as a uniform grid gives the same near-axis fields with 30 times fewer cells and 30 times less computing time. The same operators carry over to the $m=1$ mode, and the code scales across CPU cores and GPUs through pipelining along $\xi$. For the self-modulation of the AWAKE proton bunch, QUARZ agrees with HiPACE++ in the amplitude and phase of the wakefield over 40 plasma wavelengths, and 16 GPUs follow the bunch over 12~m in 5 to 11 minutes.

A fine axial mesh also exposes the singular closure of a cold-plasma bubble, which uniform grids hide by smoothing it over one cell. An explicit smoothing of the plasma sources over a length chosen independently of the grid regularizes it; the wake behind the closure then converges, and the model changes only at order of the smoothing length.

The code already contains a laser envelope solver on the same grid (a complex tridiagonal Crank--Nicolson step per slice) and field and impact ionization, which will be described separately. Natural further extensions are higher azimuthal modes and adaptive grids that follow the witness radius during the run.

Finally, QUARZ shows that a zero-touch development, in which an AI agent writes all of the code and a physicist directs it and decides, by validation against independent references, what can be trusted, can produce a production-grade simulation code in plasma physics.

%---------------------------------------------------------------------------
\begin{acknowledgments}
The QUARZ source code, its validation suite, the simulations and figures of Secs.~\ref{sec:valid}\,A--C, and a first draft of this manuscript were produced with the large language model Claude (Anthropic; model identifier \texttt{claude-opus-5-5})~\cite{Claude}, working under the direction of A.P., who did not write or edit any line of the code (zero-touch development, Sec.~\ref{sec:intro}) T.C.W. tested the code and performed the simulations and timings of Secs.~\ref{sec:awake} and \ref{sec:perf}, including the HiPACE++ runs. We thank J.~P.~Farmer for the AWAKE simulation setup. We gratefully acknowledge computing time on JUWELS and JUPITER at the J\"ulich Supercomputing Centre, granted through the project LPQED. The authors reviewed and edited the text and take full responsibility for the content of this article. Claude is not an author, because it cannot take responsibility for the work.
\end{acknowledgments}

\end{document}